\documentclass[runningheads]{llncs}
\usepackage[T1]{fontenc}
\usepackage{graphicx,verbatim}
\usepackage{lineno}
\usepackage{algorithm}
\usepackage{algpseudocode}
\usepackage{multirow}
\usepackage{amsmath}
\usepackage{amsmath}

\usepackage{amssymb}
\usepackage{color}
\begin{document}
\title{Deterministic Medical Image Translation via High-fidelity Brownian Bridges}
%

\author{Qisheng He$^{\star}$,  Nicholas Summerfield$^{\dagger\S}$, Peiyong Wang$^{\star}$, Carri Glide-Hurst$^{\dagger\S}$, 
Ming Dong$^{\star}$}  
\institute{$^{\star}$ Department of Computer Science, Wayne State University, MI, USA. \\ \email{\{Qisheng.He,pywang,mdong\}@wayne.edu} \\
$^{\dagger}$ Department of Human Oncology, University of Wisconsin-Madison, WI, USA. \\\email{glidehurst@humonc.wisc.edu} \\
$^{\S}$ Department of Medical Physics, University of Wisconsin-Madison, WI, USA. \\ \email{nsummerfield@wisc.edu}}
    
\maketitle              

\begin{abstract}
Recent studies have shown that diffusion models produce superior synthetic images when compared to Generative Adversarial Networks (GANs). However, their outputs are often non-deterministic and lack high fidelity to the ground truth due to the inherent randomness. In this paper, we propose a novel High-fidelity Brownian bridge model (HiFi-BBrg) for deterministic medical image translations. Our model comprises two distinct yet mutually beneficial mappings: a generation mapping and a reconstruction mapping. The Brownian bridge training process is guided by the fidelity loss and adversarial training in the reconstruction mapping. This ensures that translated images can be accurately reversed to their original forms, thereby achieving consistent translations with high fidelity to the ground truth. Our extensive experiments on multiple datasets show HiFi-BBrg outperforms state-of-the-art methods in multi-modal image translation and multi-image super-resolution.
\end{abstract}

\section{Introduction}
Medical image translation, the process of converting one imaging modality or representation into another, has shown incredible promise in enhancing clinical protocols. Specifically, translation has been applied to cross-modality synthesis \cite{multimodalTranslating} (synthetic CT from MRI \cite{Arabi2019} or between MRI sequences \cite{diamondGan}), super-resolution \cite{rethinkingDiffusion}, noise reduction \cite{GAnsari2020}, and reconstruction 
\cite{Kaji2019}. Generative Adversarial Networks (GANs) \cite{gan} have emerged as a promising tool in the medical image translation domain. GANs provide innovative solutions through employing two neural networks, a generator and a discriminator, that collaborate to generate high-quality synthetic images from input data. Moreover, their deterministic nature ensures consistent and reproducible results \cite{ganForCTDenoising}, a necessity for clinical reliability. 

Diffusion models \cite{ddpm} have recently gained prominence in medical image translation due to their capacity to generate high-quality images by progressively refining noisy data. In contrast to GANs, which employ adversarial training, diffusion models employ a series of denoising sampling steps to transform random noise into coherent images. This approach has demonstrated superior image quality compared to GANs \cite{diffusionBeatsGans}. More recently, Brownian bridge diffusion models (BBrg) \cite{bbdm} \cite{abridge} enhanced the diffusion process by directly learning the translation between two domains through bidirectional diffusion and achieved state-of-the-art (SOTA) sample quality among diffusion-based models. In medical applications, diffusion models have demonstrated remarkable versatility and effectiveness across various tasks, including image translation \cite{ddgan}, synthesis \cite{medsyn}, segmentation \cite{medsegdiffv2}, and 3D image generation \cite{ddpm3d}. For instance, SynDiff \cite{syndiff} employs a conditional diffusion framework with a cycle-consistent architecture to train on unpaired datasets, achieving superior image clarity and quality compared to GANs. However, despite their recent success, the non-deterministic nature of diffusion models poses a significant challenge in medical image translation \cite{adaptiveDiffusionPriors}. The variability in generated images can lead to inconsistencies, potentially undermining the reliability of diagnostic interpretations \cite{solvingDiffusionODEs}. While Ordinary Differential Equation-based sampling ensures consistency, it frequently produces blurry images, especially in diffusion bridge models, which start from a clear image rather than noise, leading to over-smoothing and loss of details \cite{ddbm}.

\begin{figure}[tbp]
    \centering
    \vspace{-0.1in}
    \includegraphics[width=0.9\linewidth]{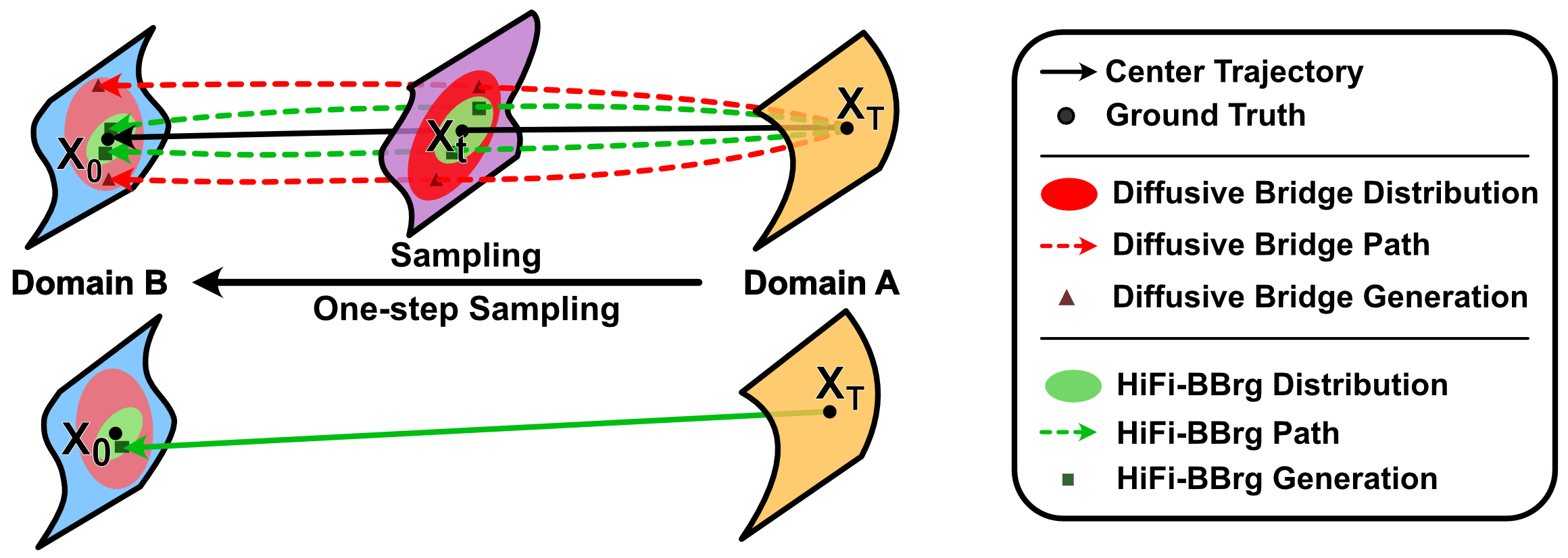}
    \vspace{-0.2in}
    \caption{Other diffusive bridge models (e.g., \cite{bbdm} and \cite{abridge}) vs. HiFi-BBrg. \textbf{Up}: Image translation from domain $\mathcal{A}$ to $\mathcal{B}$ through sampling. \textbf{Bottom}: One time-step sampling in HiFi-BBrg provides completely deterministic results.
    }
    \vspace{-0.2in}
    \label{fig.distribution}
\end{figure}

In this paper, we propose a novel High-fidelity Brownian bridge model (HiFi-BBrg) for deterministic medical image translations. Our model comprises two distinct yet mutually beneficial mappings: a generation mapping using conditional Brownian bridge diffusion model that transfers domain $\mathcal{A}$ to $\mathcal{B}$, and a reconstruction mapping using conditional GAN (cGAN) \cite{pix2pix} that transfers domain $\mathcal{B}$ to $\mathcal{A}$. The Brownian bridge training process is guided by the fidelity loss and adversarial training in the reconstruction mapping, ensuring that translated images can be accurately reversed to their original forms, thereby achieving consistent translations with high fidelity to the ground truth. The main contributions of this work are as follows: 
\begin{figure} [htbp]
    \centering
    \vspace{-0.1in}
    \includegraphics[width=0.9\linewidth]{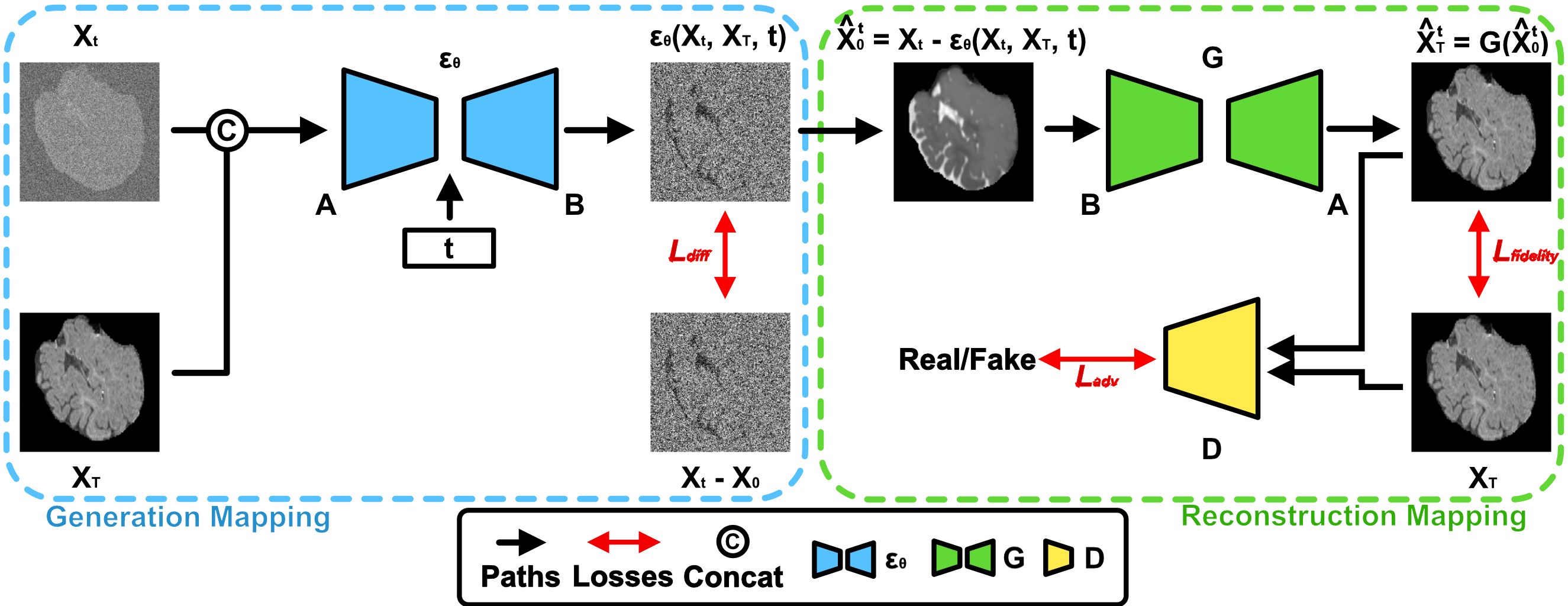}
    \vspace{-0.15in}
    \caption{HiFi-BBrg architecture illustration. The conditional Brownian bridge $\epsilon_\theta$ maps from $\mathcal{A}$ to $\mathcal{B}$, while the conditional generator $G$ maps from $\mathcal{B}$ to $\mathcal{A}$. The adversarial training is performed using the discriminator $D$ to distinguish between the real image $X_T$ and its reconstruction $\hat{X}^t_T$.}
    \vspace{-0.2in}
    \label{fig.structure}
\end{figure}
\begin{enumerate}
    \item We introduced HiFi-BBrg, a novel approach that mitigates the shortcomings of both GANs and diffusion models for medical image translation. In contrast to conventional BBrg, HiFi-BBrg generates consistent, reproducible, and high-fidelity medical images through a combined end-to-end training of a conditional Brownian bridge for generation and a cGAN for reconstruction.
    \item In HiFi-BBrg, we spread the requirement of fidelity over all steps in bridge model training by leveraging the fidelity loss and adversarial training in the reconstruction mapping. Consequently, the sampling trajectory of our model is condensed to the central one with significantly lower variance compared to other bridge models (see Figure \ref{fig.distribution}). Furthermore, our model offers completely deterministic outputs with an efficient one-step sampling process, ensuring both image quality and fidelity to the ground truth in the translated images. 
    \item Our extensive experiments on multiple datasets across various medical image translation tasks, including multi-modal image translation and multi-image super-resolution, demonstrated remarkable improvements by HiFi-BBrg when compared to SOTA methods, including both GAN-based and diffusion-based models.
\end{enumerate}

\section{High-fidelity Diffusive Brownian Bridges}
Let $\mathcal{A}$ and $\mathcal{B}$ be paired data sets of the same dimension $d$. As depicted in Figure \ref{fig.structure}, our HiFi-BBrg framework comprises two distinct yet complementary mappings: a generation mapping $\Gamma:\mathcal{A}\rightarrow \mathcal{B}$ and a reconstruction mapping $\Pi:\mathcal{B}\rightarrow \mathcal{A}$. The generation mapping employs a diffusion process, utilizing a conditional Brownian Bridge $\epsilon_\theta$ to replace random noise $\epsilon$ gradually, thereby progressively modifying the input image during sampling. The reconstruction mapping is a deterministic one implemented by cGAN via a generator $G$ and a discriminator $D$. Our algorithm aims at predicting the mapping $\Gamma$ with high fidelity, such that our model output $\hat{X}_0$ estimates the ground truth $X_0 = \Gamma(X_T)$ with high accuracy for any $X_T$ in $\mathcal{A}$.

\begin{algorithm}
\caption{Training Algorithm for HiFi-BBrg}
\label{alg.train}
\begin{algorithmic}[1]
\Function{Train}{$\epsilon_\theta$, $G$, $D$, $T$, $\lambda$}
\Repeat
    \State $X_T, X_0 \leftarrow nextBatch()$ \Comment{Get current batch from dataset}
    \State $t, \epsilon \sim \text{Uniform}(1, \ldots, T), \mathcal{N}(0, I)$ \Comment{Sampling $t$ and random noise $\epsilon$}
    \State $X_t = \frac{t}{T} X_T + (1 - \frac{t}{T}) X_0 + B(t)\epsilon$ \Comment{Forward sampling to time $t$}
    \State $L_{diff} = \| X_t - X_0  - \epsilon_\theta(X_t, X_T, t)\|$ \Comment{Diffusion loss for $\epsilon_\theta$}
    \State $\hat{X}^t_0 = X_t - \epsilon_\theta(X_t, X_T, t)$ \Comment{Calculate predictions at time $t$}
    \State $L_{cyc} = \|X_T - G(\hat{X}^t_0)\|$ \Comment{Fidelity loss}
    \State $d_f, d_r \leftarrow D(\hat{X}^t_T), D(X_T)$ \Comment{Get fake and real predictions}
    \State $L_{adv} = -(log(1-\sigma(d_f)) + log(\sigma(d_r))$ \Comment{Adversarial loss}
    \State $L = L_{diff} + \lambda L_{cyc} + L_{adv}$ \Comment{Calculate total training objective}
    \State $backwardPropagate(L)$ \Comment{Update weights with backward propagate}
\Until{converged} 
\EndFunction
\end{algorithmic}
\end{algorithm}

\subsection{Generation Mapping ($\Gamma$) via Conditional Brownian Bridge}
HiFi-BBrg adopts the Brownian bridge framework expressed in the forward stochastic differential equation (\ref{eq1}) and its reverse counterpart \cite{abridge}:
\begin{equation}\label{eq1}
dX = -\frac{X-X_T}{1-t}\,dt + 2\sqrt{1-t}\,dW(t)\ \ (0 < t < 1)
\end{equation}
where $X_T \in \mathcal{A}$ and $X(0) = X_0 \in \mathcal{B}$ are paired, and $W(t)$ is a Brownian motions. At each time $t$, $X_t$ is given by
$(1-t)X_0 + tX_T + B(t)\epsilon$ 
where $B(t) = 2(1-t)\sqrt{\ln \frac{1}{1-t}}$ is the variance of the white noise $\epsilon$. The unconditional Brownian bridge, trained with the objective $\min\|X_t - X_0 - \epsilon_{\theta}(X_t,t)\|$, yields a highly complex neural network $\epsilon_{\theta}$ that arises from estimating $X_t - X_0 = t(X_T - \Gamma(X_T)) + B(t)\epsilon$ solely based on $X_t$ and $t$. It is worth noting that incorporating $X_T$ as a condition significantly simplifies the neural network $\epsilon_{\theta}$ by enabling the estimation of $X_t - X_0$ using $X_t$, $X_T$, and $t$. Accordingly, the training objective of this conditional Brownian bridge, denoted as diffusion loss, is given by:
\begin{equation}
\label{eq.objective}
L_{diff} = \|X_t - X_0 - \epsilon_{\theta}(X_t,X_T,t)\|
\end{equation}

\begin{table}
    \centering
    \begin{tabular}{c | c | c c c c}
        \hline
        \multirow{2}{*}{\textbf{Model}} & \textbf{Sampling} & \multirow{2}{*}{\textbf{LPIPS} $\downarrow$} & \multirow{2}{*}{\textbf{PSNR} $\uparrow$} & \multirow{2}{*}{\textbf{SSIM} $\uparrow$} & \multirow{2}{*}{\textbf{Std.} $\downarrow$} \\
        & \textbf{Steps} \\
        \hline
        BBrg & 1000 & 0.2260 & 22.2 & 0.783 & 0.0285 \\
        Conditional BBrg & 1000 & 0.1126 & 25.7 & 0.641 & 0.0078 \\
        \hline
        HiFi-BBrg & 1000 & 0.0953 & 25.2 & \textbf{0.909} & 0.0021 \\
        HiFi-BBrg & 200 & 0.0951 & 25.5 & \textbf{0.909} & 0.0002 \\
        HiFi-BBrg & \textbf{1} & \textbf{0.0860} & \textbf{26.6} & 0.908 & \textbf{0.0000} \\
        \hline
    \end{tabular}
    \caption{Quantitative scores with different components and sampling steps in HiFi-BBrg on iSEG 2017 (T1-W to T2-W modality translation). Note that with unconditional/conditional generation mapping and no reconstruction mapping, HiFi-BBrg is reduced to BBrg/Conditional BBrg, respectively. The best results are \textbf{bolded}, and the same applies for the remaining tables.}
    \vspace{-0.2in}
    \label{table.ablation.components}
\end{table}

\begin{table}
    \centering
    \begin{tabular}{c | c c c c}
        \hline
        \textbf{Training Time Steps} & \textbf{LPIPS} $\downarrow$ & 
       \textbf{PSNR} $\uparrow$ & \textbf{SSIM} $\uparrow$ \\
        \hline
        1 & 0.1320 & 24.5 & 0.858 \\
        200 & 0.1164 & 26.2 & 0.873 \\
        1000 & \textbf{0.0860} & \textbf{26.6} & \textbf{0.908} \\
        \hline
    \end{tabular}
    \caption{Quantitative scores with different time steps during training in HiFi-BBrg using one time step sampling on iSEG 2017 (T1-W to T2-W modality translation).}
    \vspace{-0.3in}
    \label{table.ablation.time}
\end{table}

\subsection{Reconstruction Mapping ($\Pi$) via cGAN}
While the training objective in Eq. \ref{eq.objective} of the Brownian bridge generates images of high quality, we additionally need to ensure the sample fidelity. For this purpose, at every time step, we estimate $X_0$ via $\hat{X}^t_0 = X_t - \epsilon_\theta(X_t, X_T, t)$ and generate an estimate of $X_T$ via a cGAN: $\hat{X}^t_T = G(\hat{X}_0)$. By comparing the reconstructed $\hat{X}^t_T$ to the input $X_T$, we spread the requirement of fidelity over all steps just as we train the Brownian bridge objective:
\begin{equation}
    \label{eq.fidelity}
    L_{fidelity} = \|X_T - \hat{X}^{t}_T\|
\end{equation}
Consequently, the sampling trajectory of HiFi-BBrg is condensed to the center one with a much lower variance compared with other bridge models.

The adversarial training objective of the cGAN can be formulated as:
\begin{equation}
    \label{eq.adv_loss}
    L_{adv} = \mathbb{E}_{\hat{X}^t_T \sim p(\hat{X}^t_T)}[log(1-\sigma(d_f))] + \mathbb{E}_{x \sim p(x_T)}[log(\sigma(d_r))]
\end{equation}
where $\sigma(\cdot)$ denotes the sigmoid function, and $d_f$ and $d_r$ denote the fake and real logits generated by the discriminator, respectively. This objective is integrated into the cGAN training process. 

The final training objective merges this adversarial loss with the diffusion and fidelity loss, directing the model to produce high quality and fidelity samples:
\begin{equation}
    \label{eq.total_loss}
    L = L_{diff} + \lambda L_{fidelity} + L_{adv}
\end{equation}
where $\lambda$ is a constant coefficient to balance the fidelity loss. Finally, we directly predict $\hat{X}^T_0$ via a deterministic one-step sampling as $X_T - \epsilon_\theta(X_T, X_T, T)$. 
Therefore, HiFi-BBrg provides completely deterministic results with an efficient one-step sampling, ensuring both image quality and fidelity to the ground truth in the translated medical images. The comprehensive HiFi-BBrg training algorithm is presented in Algorithm \ref{alg.train}.


\begin{table}
    \centering
    \begin{tabular}{c | c c c}
        \hline
        \textbf{Model} & \textbf{LPIPS} $\downarrow$ & 
        \textbf{PSNR} $\uparrow$ & \textbf{SSIM} $\uparrow$ \\
        \hline
        Pix2Pix \cite{pix2pix} & 0.1132 & 21.1 & 0.82 \\
        CycleGAN \cite{cyclegan} & 0.0605 & 23.5 & 0.83 \\
        RegGAN \cite{reggan} & 0.0605 & 25.9 & 0.86 \\
        DDPM \cite{ddpm} & - & 26.3 & 0.89 \\
        Fast-DDPM$^*$ \cite{fastddpm} & 0.1038 & 26.1 & 0.91 \\
        \hline
        \textbf{Ours} & \textbf{0.0401} & \textbf{29.9} & \textbf{0.94} \\
        \hline
    \end{tabular}
    \caption{Quantitative comparison on the BraTS 2018 dataset, transferring from T1-W to T2-W Modality. Methods marked with * represent the mean of five sampling trials.}
    \vspace{-0.2in}
    \label{table.brats}
\end{table}

\begin{table}
    \centering
    \begin{tabular}{c | c c}
        \hline
        \textbf{Method} & \textbf{PSNR} $\uparrow$ & \textbf{SSIM} $\uparrow$ \\
        \hline
        miSRCNN \cite{misrcnn} & 26.5 & 0.87 \\
        miSRGAN \cite{misrgan} & 26.8 & \textbf{0.88} \\
        DDPM \cite{ddpm} & 25.3 & 0.83 \\
        Fast-DDPM$^*$ \cite{fastddpm} & 26.7 & \textbf{0.88} \\
        \hline
        Ours & \textbf{31.2} & \textbf{0.88} \\
        \hline
    \end{tabular}
    \caption{Quantitative comparison on the Prostate MRI dataset for image super-resolution. Again, methods marked with * represent the mean of five sampling trials.}
    \vspace{-0.3in}
    \label{table.prostateMRI}
\end{table}

\section{Experiments}
\subsection{Datasets}
For quantitative and qualitative evaluation of our model, we selected three distinct datasets for the following tasks: \textbf{iSEG 2017} \cite{iseg} for ablation studies, \textbf{BraTS 2018} for multi-modal image translation \cite{brats}, and \textbf{Prostate MRI} \cite{prostatemri} for multi-image super-resolution. All images in these datasets are resized to $256 \times 256$ and normalized to the range $[-1, 1]$, with the exception of BraTS 2018, where images were padded using the method described in \cite{reggan}.

\subsection{Metrics and Implementation Details}
We adopted the Brownian bridge \cite{abridge} as $\epsilon_\theta$, concatenating its condition into the channel dimension of the input image $X_t$ at each time step $t$. We employed the same UNet architecture without time embeddings for the reconstruction generator $G$. Furthermore, we adopted the discriminator $D$ from \cite{pix2pix}.

Empirically, we set $\lambda$ to 1 for the fidelity loss for BraTS 2018 and Prostate MRI, and 0.5 for the iSEG 2017 dataset. All models were trained for 1000 epochs using the Adam optimizer. The learning rate was set to $2 \times 10^{-5}$ for $\epsilon_\theta$ and $G$ following \cite{fastddpm}, and $2 \times 10^{-4}$ for $D$ following \cite{pix2pix}.

To assess the quality of the translated images and the fidelity between the generated and the ground truth images, we employed three metrics: Learned Perceptual Image Patch Similarity (LPIPS) \cite{lpips} to evaluate perceptual realism, peak signal-to-noise ratio (PSNR) to measure pixel-wise differences, and the structural similarity index measure (SSIM) \cite{ssim} to capture structural accuracy.

All our experiments were conducted using the PyTorch \cite{pytorch} and Monai \cite{monai} frameworks and were executed on a machine equipped with two NVIDIA RTX A40 GPUs. The source code for this work will be publicly released after the review period of MICCAI 2025.

\subsection{Ablation Study}
We first conducted a series of ablation studies to systematically assess different components and sampling steps in HiFi-BBrg on the iSEG 2017 dataset for T1-W to T2-W modality translation. 

\begin{figure}
    \centering
    \vspace{-0.1in}
    \includegraphics[width=0.8\linewidth]{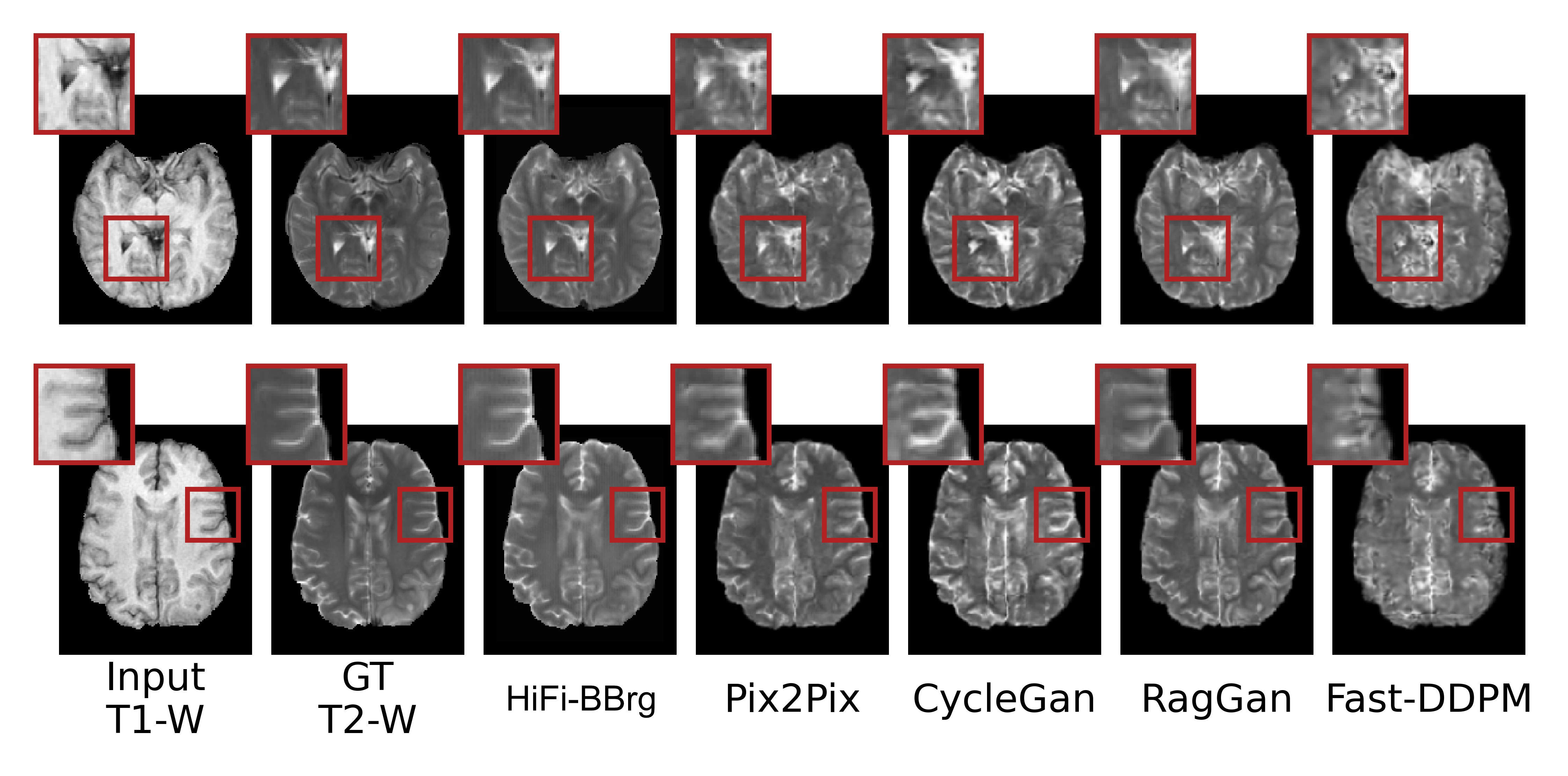}
    \vspace{-0.15in}
    \caption{Comparison of HiFi-BBrg and SOTA Methods on the BraTS2018 T1-W to T2-W Image Translation Dataset. The images generated by Fast-DDPM method were sampled five times, resulting in a std. of 0.0072.}
    \vspace{-0.2in}
    \label{fig.BraTS2018}
\end{figure}

The results presented in Table \ref{table.ablation.components} clearly demonstrate that BBrg generated non-realistic and inaccurate images due to its unconstrained sampling process. As a comparison, GAN-based Pix2pix \cite{pix2pix} achieved a result of 0.1675 LPIPS, 24.5 PSNR, and 0.869 SSIM. By incorporating a condition, conditional BBrg generated more realistic images while exhibiting significantly lower pixel-level differences. Guided by the fidelity loss and adversarial training in the reconstruction mapping, the images generated with the same 1000 sampling steps by HiFi-BBrg exhibited higher quality and fidelity to the ground truth. Furthermore, we evaluated the performance of HiFi-BBrg using various sampling time intervals. HiFi-BBrg achieved superior LPIPS and PSNR when employing a single sampling step, suggesting that the samples exhibited greater realism and were more closely aligned with the ground truth. Notably, the results are entirely deterministic, as no noise is introduced during the process.

Secondly, we conducted an ablation study on HiFi-BBrg by varying the training time step $T$. The trained models were subsequently evaluated using a single time step sampling. As presented in Table \ref{table.ablation.time}, HiFi-BBrg models trained with more time steps exhibited superior image quality and fidelity. Based on these ablation results, HiFi-BBrg was configured with 1000 time steps during training and a single sampling step for deterministic and highly efficient sampling during image translation in the rest of our experiments.




\subsection{Quantitative Evaluation}
Next, we conduct a quantitative evaluation of our HiFi-BBrg model. We selected Pix2Pix \cite{pix2pix}, CycleGAN \cite{cyclegan}, RegGAN \cite{reggan}, DDPM \cite{ddpm}, and Fast-DDPM \cite{fastddpm} for comparison on BraTS 2018 (multi-modal image translation); and miSRCNN \cite{misrcnn}, miSRGAN \cite{misrgan}, DDPM \cite{ddpm}, and Fast-DDPM \cite{fastddpm} for comparison on the Prostate MRI (multi-image super-resolution). 

\begin{figure}
    \centering
    \vspace{-0.1in}
    \includegraphics[width=0.8\linewidth]{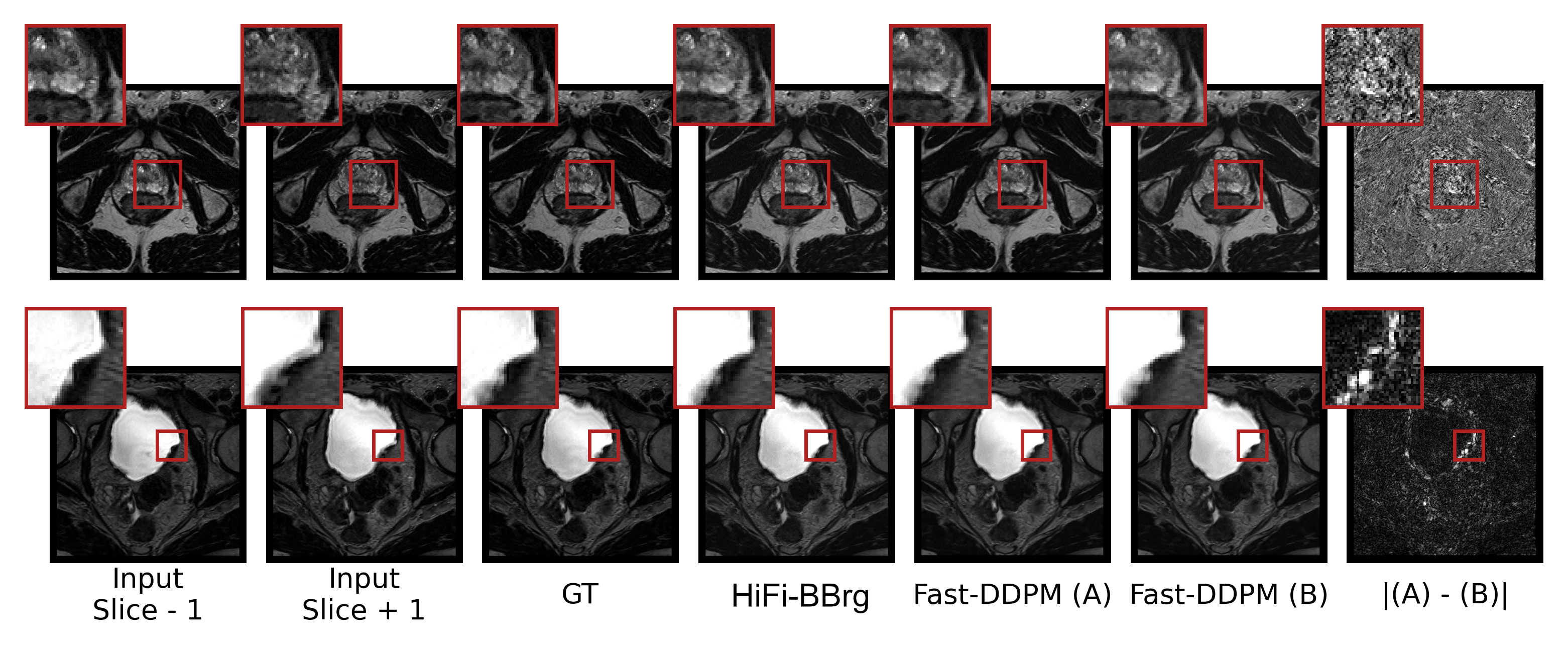}
    \vspace{-0.15in}
    \caption{Comparison between HiFi-BBrg and SOTA methods on the Prostate MRI super-resolution dataset. The images generated by Fast-DDPM method were sampled five times, resulting in a std. of 0.0104. Two samples are shown with an absolute difference plot highlighting variations due to the non-deterministic approach.}
    \vspace{-0.2in}
    \label{fig.prostate}
\end{figure}

Table \ref{table.brats} presents a comparative analysis of the performance of HiFi-BBrg with other SOTA methods on BraTS 2018 for T1-W to T2-W modality translation. Notably, HiFi-BBrg significantly surpasses both GAN-based and diffusion-based models across all three metrics — LPIPS, PSNR, and SSIM — demonstrating its superior image quality and high fidelity to the ground truth.

Table \ref{table.prostateMRI} presents a comparative analysis of the performance of HiFi-BBrg on the Prostate MRI dataset for image super-resolution. HiFi-BBrg outperforms the SOTA methods in terms of PSNR, attaining a value of 31.2 with a significant improvement of +4.5 over the second best. It also achieves a tied best SSIM value of 0.88. Note that HiFi-BBrg delivers superior results than specialized image super-resolution algorithms such as miSRGAN \cite{misrgan} even though it is a general image-to-image translation model.

\subsection{Qualitative Evaluation}
Additionally, we conduct a qualitative evaluation of the image quality of HiFi-BBrg. First, axial slices of the brain are presented in Figure \ref{fig.BraTS2018} for BraTS 2018. The HiFi-BBrg method accurately preserves the key anatomical structures, such as ridges, folds, and ventricles, present in the source domain. HiFi-BBrg generates visual textures that are comparable to those of the target domain while maintaining tissue boundaries. In contrast, other methods appear to capture global characteristics but fail to reliably represent the same tissues locally. Their predictions lose important structural information of the brain folds while resulting in an overall more noisy appearance.

The Prostate MRI super-resolution dataset is demonstrated in Figure \ref{fig.prostate} with a pelvic axial slice through the prostate and bladder. Fast-DDPM was sampled twice on the same input, and an absolute difference plot is provided to highlight its non-deterministic nature. Overall, both methods successfully capture fine details, with tissue texture closely matching that of the reference slice. However, Fast-DDPM introduces minor local variations, as highlighted in the zoomed-in regions, whereas our HiFi-BBrg's deterministic predictions remain consistent. 

\section{Conclusion and Future Work}
In this paper, we propose HiFi-BBrg for deterministic medical image translations. Our extensive experiments on diverse datasets demonstrate that HiFi-BBrg outperforms SOTA methods in multiple medical translation tasks. In the future, we plan to extend  HiFi-BBrg to handle unpaired training data by introducing an architecture to couple two HiFi-BBrg modules that bilaterally translate between different domains.

\clearpage

\small
\bibliographystyle{ieeetr}
\bibliography{references}

\end{document}